\documentstyle[12pt,aaspp4]{article}  
  
\begin{document}  
 
\begin {flushright} 
OU-TAP-53 \\ 
January 1997 
\end{flushright} 
 
\title{\bf STABILITY OF SCALE-INVARIANT COSMOLOGICAL 
CORRELATION FUNCTIONS IN THE STRONGLY NON-LINEAR CLUSTERING REGIME}  
  
\author{Taihei Yano and Naoteru Gouda}  
  
\affil{Department of Earth and Space Science,  
Graduate School of Science, Osaka University  
Toyonaka, Osaka 560, Japan\\  
E-mail: yano, gouda@vega.ess.sci.osaka-u.ac.jp}  
  
\begin{abstract}  
We investigate stability of the scale-invariant solutions of the BBGKY 
equations for two-point spatial correlation functions of the density 
fluctuations in the strongly non-linear regime.
In the case that the background skewness of the velocity field is equal to 0, 
we found that there is no local instability in the strongly non-linear regime.
The perturbation does not grow nor does it decay.
It has an only marginal stable mode.
This result means that no special value of the power index of the two-point 
spatial correlation function are favored in terms of the stability of the 
solutions. In other words, the argument about the stability does not 
determine the power index of the two-point spatial correlation functions 
in the strongly non-linear regime.
\end{abstract}  

{\it Subject headings}:  
cosmology:theory-large scale structures-correlation function   
  
\section{INTRODUCTION}  

The large scale structure formation is one of the most important probrems in 
the cosmology. It is generally believed that these structures have 
been formed due to gravitational instability.
 Hence it is very important to clarify the evolution of density fluctuations  
by gravitational instability.
Here we consider the density fluctuations of the collisionless particles 
such as dark matters because our interest is concentrated on the effect of 
the self-gravity.  

 When the density fluctuations are much smaller than unity,
that is, in the linear regime, 
time evolutions of the small density fluctuations   
can be analyzed by making use of the linear theory.  
  In this regime, we can understand analytically how the small fluctuations   
grow(Peebles 1980, 1993).  
But  when the density fluctuations are much larger than unity,
that is, in the strongly nonlinear regime, 
the analytical approach is very difficult.
But we believe that it is very necessary to understand clearly 
the nonlinear behavior of the density fluctuations  
on these strongly nonlinear scales
because galaxy formations are much related to the density fluctuation 
on these small scales and also it is a very interesting academic problem
for the nonlinear dynamics of the self-gravity.

As one way to quantify the clustering pattern, we generally use 
a two-point spatial 
correlation function of density fluctuations. In the strongly non-linear 
regime, it is found   
{}from $N$-body simulations that two-point spatial correlation 
functions obey the power law.  This result is reasonable because the 
self-gravity is scale-free.  
 Two-point spatial correlation function
has been investigated by various methods
because the power index of the two-point spatial correlation 
function is a good  
indicator representing the nonlinear dynamics of the self-gravity in this  
regime. 
  
The power index of the two-point spatial correlation function have been   
usually analyzed by   $N$-body simulations(Frenk, White, \& Davis 1983; 
Davis et.al. 1985; Suto 1993 and references therein).
However, the physical process that determine the value of the power 
index cannot be clarrified only by the $N$-body simulations.  
There are some other methods besides this numerical simulation.  
One of them is the analysis by the BBGKY equations. The work by Davis \&  
Peebles (1977; hereafter DP) is a pioneer for the analysis by the BBGKY 
equations.  They showed the existence of the self-similar solutions for 
spatial correlation functions under some assumptions.  Then it is shown that 
the power index $\gamma$ of the two-point spatial correlation function in the 
strongly nonlinear regime is related to the initial power index $n$ of the 
initial power spectrum $P(k)$ as follows;  
  
\begin{equation}  
\xi(r)\propto   
               r^{-\gamma}  (\xi\gg 1~:~\gamma=  
               \frac{\displaystyle 3(3+n)}{\displaystyle 5+n})  
\label{1}
\end{equation}  
  
One of the assumptions that DP adopted is called the stable condition.  
This condition is that the mean relative physical velocity in the 
strongly nonlinear regime is equal to zero. 
This condition was tested by the $N$-body simulations  
(Efstathiou et al. 1988; Jain 1995), but this condition is not 
completely verified.  Furthermore the stability of the 
scale-invariant solutions in the strongly nonlinear regime, 
which DP derived, was investigated by the linear perturbation 
theory(Ruamsuwan \& Fry 1992;hereafter RF)
because there is no guarantee that such solutions can exist stably.
And it is found 
that the solutions are marginally stable.  
  
As for the physical process   
determining the power index in the strongly nonlinear regime, there are other 
analysis besides one proposed by DP.  One of them is given by Saslaw(1980).  
He concluded that the power index $\gamma$ approaches to 2 by using the 
cosmic energy equation under some assumptions while some numerical 
simulations do not support this result
(Frenk, White \& Davis 1983; Davis et al. 1985; Fry \& Melott 1985).    
  
There is another idea as follows; when the initial power spectrum has the 
sharp cut-off or the initial power spectrum is scale-free with negative and 
small initial power index, then there appear anywhere caustics of the density 
fields.  In these cases, the power index is irrespective of the detailed 
initial conditions after the first appearance of caustics on the small scales 
around the typical size of the thickness of caustics(in three-dimensional 
systems, they correspond to the pancake structures of highly clustered 
matters).  The power index is determined by the type of these caustics 
which is classified in accord with the catastrophe theory.  
This idea is verified in the one-dimensional system
(Kotok \& Shandarin 1988; Gouda \& Nakamura 1988, 1989), the spherically 
symmetric systems(Gouda 1989), the two-dimensional systems (Gouda 1996a) 
and also the three-dimensional systems(Gouda 1996b).  
In these cases, it is suggested that 
$\gamma \approx 0$ on the small scales.    
As we can see from the above arguments, we believe that there are still 
uncertainties about the physical processes which determine the value of 
the power index.

 Yano \& Gouda (1996;hereafter YG) investigated the conditions that 
determine the power index of the two-point spatial correlation function
in the strongly nonlinear regime by analysing the scale-invariant solutions 
of the BBGKY equations in this regime.
YG does not adopt the following assumptions that DP adopted; (1) the 
skewness is equal to 0, (2) three-point 
spatial correlation function is represented by a product of the two-point 
spatial correlation function, (3) the stable condition is satisfied.
As a result, YG obtained the relation between a mean relative pecurier 
velocity, skewness, three-body correlation function and the power index 
of the two-point spatial correlation function.  
YG found that the stable condition is not independent one of the other 
assumptions (1) and (2). That is, the stable condition is satisfied only
when both assumptions (1) and (2) are satisfied. The assumptions (1) and (2)
cannot be generally satisfied in all cases and then there is no guarantee 
that the stable condition are correct. Furthermore YG suggests from the 
physical point of view the probable range of the mean peculier velocity 
which includes the the value given by the stable condition.
This fact results in the possibility that the power index of the 
two-point spatial correlation functions takes various values according to the 
mean peculier velocity.
Indeed, YG showed that the mean relative physical pecurier velocity 
can take the value between 0 (stable clustering) and the Hubble expansion 
velocity (comoving clustering).
When the stable clustering picture and the self-similerity are satisfied,
the power index of the two-point spatial correlation function have 
the value that DP derived.
On the other hand, when the comoving clustering picture and 
the self-similerity are satisfied,
the power index of the two-point spatial correlation function have 
the value of 0. And this value is consistent with the result from 
the catastrophy theory(Gouda\& Nakamura 1988 1989;Gouda 1989 1996a;b).
Although we found that there exist various scale-invariant solutions
with different power index,
whether these solutions 
are stable or not is an another interesting probrem.
Some values of the power index of the two point spatial correlation 
function may not be able to be taken in the real world 
if these solutions are unstable. 
 RF investigated the stability of the DP solutions 
by making use of the linear perturbation.
 And they showed that the perturbations of the solutions is 
marginal stable. 
But they investigated only in the DP case and did not investigate the 
other solutions that YG obtained. Furthermore, as we will discuss later,
RF mistook the way of providing the perturbation about the skewness
although their result is correct fortunately due to the reason 
which we will show later ($\S 3$).  
Furthermore they did not comment about the 'strange' growing mode 
which exist in their solutions of the perturbation.
We will show that this 'strange' modes are resulted from the  fact that 
RF put the inapplicable form of the perturbation which diverges on small 
or large scales. 
So, in this paper, we investigate the stability of the general solutions 
that YG obtained by putting the applicable form of the perturbation.
We will derive the perturbation equation by perturbed BBGKY equation 
from the background scale-invariant solutions in the strongly 
nonlinear regime. And we will show the solution of the linear 
perturbation equation when we assume the appropriate form of the 
perturbation which is well-defined.

In $\S 2$, we briefly show the BBGKY equations that we use in this paper
and also show the scale-invariant solutions in the strongly non-linear 
regime that YG obtained.
In $\S 3$, we consider the stability of these solutions 
of the BBGKY equations by the analysis of the linear perturbation.  
 Finally, we devote $\S 4$  to conclusions and discussions.

\section{BASIC EQUATIONS AND THE SCALE-INVARIANT SOLUTIONS}
  
At first, we show the BBGKY equations that we use in this paper.
These equations are time evolution equations of the statistical value 
such as two-body correlation function, three-body correlation function,
and so on.
These equations can be derived from the ensumble mean of the Vlasov 
equation (DP,RF).  
The $N$-th BBGKY equation represents the time evolution 
of the $N$-body correlation 
function. We are now interested in the two-body correlation function,
and then, we use the second BBGKY equation by taking the momentum moment of 
this equation.
Zeroth moment of the second BBGKY equation becomes the time evolution 
equation of the two-point spatial correlation function. 
This equation involves the first moment term, that is, the term of the 
 mean relative peculier velocity.
The first moment of the second BBGKY equation becomes the time evolution
equation of the mean relative peculier velocity.
This equation involves the second moment term of the relative peculier 
velocity dispersion $\Pi$, $\Sigma$,
where, $\Pi$ ,and $\Sigma$ are pararel and transverse component 
of the mean relative peculier velocity dispersion, respectively (DP,RF,YG). 
The second moment of the second BBGKY equation becomes the time evolution
equations of the relative peculier velocity dispersion.
These equations invole the skewness of the velocity field in the same way.
In deriving the BBGKY equations, DP assumed that the skewness is equal to
0, and also the stable condition in which the mean relative peculier physical 
velocity is equal to 0 ($\langle v \rangle =-\dot{a}x$).
Furthermore DP assumed that a three-point spatial correlation function 
can be represented by 
a product of the two-point spatial correlation function as follows;

\begin{equation}  
\zeta_{123} = Q(\xi_{12} \xi_{23} +\xi_{23} \xi_{31} +\xi_{31} \xi_{12}).
\label{2}
\end{equation}  

On the other hand, RF incorporated the skewness in the equations.
RF expressed the skewness by using the values  $A$, and $B$.
These are related to our form by the next relations.

\begin{equation}  
A(3v\Pi_{RF}+v^3)=3\langle v \rangle \Pi + s_\parallel,~~~~~~~~~~
Bv\Sigma=  \langle v \rangle \Sigma + s_\perp,
\end{equation}  
where $v$ in RF is the same as $\langle v \rangle$ in our paper,
 and $\Sigma$ in RF is also the same as our $\Sigma$.
The definition of $\Pi_{RF}$ is different from ours 
($\Pi_{RF}+v^2=\Pi$). Here $s_\parallel$, $s_\perp$ are, as we will 
define later, the parallel component and the transverse component 
of the skewness, respectively.
When we investigate the scale-invariant solutions of the BBGKY equations, 
the difference of the definition of the valuables in RF form and ours
is not important. But when we perturbe
the each valuable, the skewness also must be perturbed independently.
However it must be noted that
 RF treated $A$ and $B$ as constant values.
Then RF did not correctly treat with the perturbation of the skewness.
RF used the same assumption that DP used about the three-point 
spatial correlation function.
Furthermore, they also used the stable condition. 
We do not know whether the assumption of the three-point 
spatial correlation function is correct or not.
Then we assume the three-point spatial correlation function by the following
 form;

\begin{equation}  
\zeta_{123} = Q(\xi_{12}^{1+\delta} \xi_{23}^{1+\delta} 
               +\xi_{23}^{1+\delta} \xi_{31}^{1+\delta}
               +\xi_{31}^{1+\delta} \xi_{12}^{1+\delta}),
\end{equation}  
where $\delta$ is a constant value. This is not a general form,
but an extention of the form that DP adopted. 
We use this form as a preliminary step for our analysis. 
Furthermore the stable 
condition is a special case of the mean relative peculier velocity.
Hence we do not assume the stable condition and consider the general
condition as shown in YG.
We are interested in the strongly nonlinear regime,
and then we take the non-linear approximation.
In the strongly nonlinear regime ( $x \ll 1$),
 the two-point spatial correlation function is much   
larger than unity, $\xi \gg 1$. 
In this limit,
we obtain the following four equations (DP 1977;RF 1992;YG 1996);
\begin{equation}  
    \frac{\partial \xi}{\partial t}  
+   \frac{1}{a}\frac{1}{x^2}  
\frac{\partial}{\partial x}[x^2 \xi \langle v \rangle]=0,  
~~~~~~~~~~~~({\rm 0th~moment})  
\label{12}  
\end{equation}  
\begin{eqnarray}  
       \frac{1}{a x^2}
       \frac{\partial}{\partial x}(x^2 \xi \Pi)  
&-&    \frac{ 2 \xi \Sigma}{a x}
                         \nonumber \\  
&+&  2Gm\bar{n}aQ 
\frac{x^\beta}{x}
\int
\frac{x^\beta _{31} }{x_{31}^3}
\{ \xi(x)^{1+\delta}+\xi(z)^{1+\delta}\}\xi(z-x)^{1+\delta}d^3 x_3=0,
   \nonumber \\  
&&~~~~~~~~~~~~~~~~~~~
~~~~~~~~~~~~~~~~~~~({\rm 1st~moment})  
\label{13}  
\end{eqnarray}  
\begin{eqnarray}  
\lefteqn{
         \frac{1}{a^2}\frac{\partial}{\partial t} [a^2 \xi \Pi]
+        \frac{1}{a x^2}  
         \frac{\partial}{\partial x}  
         \left[  
          x^2 \xi \{3\langle v \rangle\Pi+s_\parallel \}  
         \right]    
 -       \frac{4\xi}{a x} \{\langle v \rangle \Sigma   + s_\perp  \}  
         }  \nonumber \\
&+&  4Gm\bar{n}aQ^* 
\frac{x^\beta x^\gamma}{x^2}
\int 
\frac{x^\beta _{31} }{x_{31}^3}
\{ \xi(x)^{1+\delta}+\xi(z)^{1+\delta}\}\xi(z-x)^{1+\delta}
\langle v^\gamma \rangle d^3 x_3=0,
                                   \nonumber \\  
&&~~~~~~~~~~~~~~~~~~~~~~~  
~~~~~~~~~~~~~({\rm 2nd~moment:contraction~ 1})  
\label{14}  
\end{eqnarray}  
\begin{eqnarray}  
      \frac{1}{a^2}\frac{\partial}{\partial t}  
      [a^2 \xi \Sigma]   
&+&   \frac{1}{a}\frac{1}{x^4}  
      \frac{\partial}{\partial x}  
          \left[  
           x^4 \xi \{ \langle v \rangle \Sigma + s_\perp \}  
           \right] = 0,  
                                    \nonumber \\  
&&~~~~~~~~~~~~~~  
~~~~~~~~~~~({\rm 2nd~moment:contraction~2})  
\label{15}  
\end{eqnarray}  
 where $G$ is the  gravitational constant, $m$ is the mass of a particle,
$\bar{n}$ is the mean number density of the particles, 
$Q^*$ is the coeficient of the first momentum moment of the three-body
correlation function (RF,YG).
As YG commented, the fourth term of eq.(\ref{14}) is not satisfied in 
general although this term is correct in the strongly non-linear regime 
in general. However DP showed the existence of 
the three-body correlation function which gives this term.
As commented later, the condition that  $Q^* =Q$ is consistent 
with the zero skewness case [see eq.(\ref{45})]. 
Since we consider the case of the zero skewness, 
we assume the fourth term of eq.(\ref{14}) is satisfied in our analysis.

We express the skewness by the following expression.
  
\begin{equation}  
 s^{\alpha \beta \gamma}  
\equiv  
\langle(v-\langle v \rangle)^\alpha
       (v-\langle v \rangle)^\beta 
       (v-\langle v \rangle)^\gamma \rangle  
=s_\parallel ~ P_{ppp}^{\alpha \beta\gamma}  
 + s_\perp ~ P_{ptt}^{\alpha \beta \gamma},
\end{equation}  
  
\begin{equation}  
P_{ppp}^{\alpha \beta\gamma}  
=\frac{x^\alpha x^\beta x^\gamma}{x^3},~~~~  
 P_{ptt}^{\alpha \beta \gamma}  
=  \frac{x^\alpha}{x}\delta^{\beta \gamma}  
  +\frac{x^\beta}{x}\delta^{\gamma \alpha}  
  +\frac{x^\gamma}{x}\delta^{\alpha \beta}   
  -3\frac{x^\alpha x^\beta x^\gamma}{x^3}.  
\end{equation}  
where the subscripts $p$ and $t$ represent the parallel and transverse 
component of each two particles, respectively.   
Here $P_{ppt}$ and $ P_{ttt}$ vanish because of the symmetry of   
the background universe.  
We use the Einstein-de Sitter Universe through this paper 
because we are interested only in the scale-invariant correlations.
  
Here we consider the scale-invariant solutions of these equations.  
In the strongly nonlinear regime, it is naturally expected that the effect of 
the nonlinear gravitational clustering dominates and then the solutions in 
this regime have no characteristic scales, that is, they are expected to obey 
the power law due to the scale-free of the gravity. Then we investigate the 
power law solutions of the $\xi, \langle v \rangle, \Pi$ and $\Sigma$ (YG).
We assume that the two-point spatial correlation function $\xi$ is given by  
\begin{equation}  
\xi = \xi_0 a^\beta x^{-\gamma}.  
\end{equation}  
  
Then, we obtain from the dimensional analysis in eq.(\ref{12})

\begin{eqnarray}  
\langle v \rangle &=& -h\dot{a}x,   \nonumber \\  
\beta &=& (3-\gamma )h.
\end{eqnarray}  

In this case, the solutions of the other valuables are given by
  
\begin{eqnarray}  
\Pi&=&\Pi_0 a^{\beta(1+2\delta)-1}x^{2-\gamma(1+2\delta)}, \nonumber \\  
\Sigma&=&\Sigma_0 a^{\beta(1+2\delta)-1}x^{2-\gamma(1+2\delta)}, \nonumber \\  
s_\parallel&=&s_{\parallel_0}\dot{a}  
 a^{\beta(1+2\delta)-1}x^{3-\gamma(1+2\delta)},\nonumber \\    
s_\perp&=&s_{\perp_0}\dot{a}  
 a^{\beta(1+2\delta)-1}x^{3-\gamma(1+2\delta)}.   
\end{eqnarray}  
  
{}From eq.(\ref{15}), it is found that  
\begin{equation}  
2\beta(1+\delta)+1-\{7-2\gamma(1+\delta)\}(h-\Delta) = 0,  
\end{equation}  
and then  
\begin{equation}  
h=\frac{1+\{7-2\gamma(1+\delta)\}\Delta}{1-6\delta}   
                                    \label{20},  
\end{equation}

\begin{equation}  
\Delta \equiv \frac{s_{\perp0}}{\Sigma_0}.  
\end{equation}  

As we can see from eq.(\ref{20}), the parameter $h$ can take various values 
according to the skewness $\Delta$, the power index $\gamma$ and $\delta$.
Only when $\Delta=\delta=0$  is satisfied, the stable condition ($h=1$) is
correct.
If the similarity solutions exist, the power index of the 
two-point spatial correlation function can be represented by the following form
(Padmanabhan 1995;YG 1996);
   
\begin{equation}  
\gamma = \frac{3(3+n)h}{2+(3+n)h}.
\end{equation}  
Then, even when we assume that the self-similarity solutions exist, the power 
index of the two-point spatial correltion function can take various values 
according to the mean relative peculier velocity, $h$
(if the stable condition is satisfied, i.e., $h=1$, the result of 
DP[eq.(\ref{1})] is reproduced). 

\section{LINEAR STABILITY OF SCALE-INVARIANT SOLUTION}  

In the previous section, we showed that 
there are various scale-invariant solutions in addition to the solutions 
that DP obtained (YG 1996).
But there is no guarantee that all the solutions can exist stably.
As RF did, we investigate the stability of the solutions in the 
strongly nonlinear regime by making use of the linear perturbations theory. 
Now we  perturbe the two-point spatial correlation function from 
the scale-invariant solution
that we obtained in the previous section as follows;

\begin{equation}  
\xi '=\xi(1+\Delta_{\xi}),
\end{equation}  
where $\xi$ is the scale-invariant solution, $\xi '$ is the perturbed one and
$\Delta_\xi \ll 1$.
We also perturbe the other valuables such as the mean relative 
pecurier velocity $\langle v \rangle $, the relative pecurier velocity 
dispersions $\Pi$, $\Sigma$ and the 
skewness $s_\parallel$, $s_\perp$ in the same way.

The equations of the linear perturbations are following;
 
\begin{equation}  
    \frac{\partial \xi \Delta_\xi}{\partial t}  
+   \frac{1}{a x^2}  
\frac{\partial}{\partial x}[x^2 \xi \langle v \rangle 
\{\Delta_\xi +\Delta_{\langle v \rangle}\}]=0,
~~~~~~({\rm 0th~moment})  
\label{23}  
\end{equation}  
\begin{eqnarray}  
       \frac{1}{a x^2}
       \frac{\partial}{\partial x}[x^2 \xi \Pi\{\Delta_\xi +\Delta_{\Pi}\}]  
&-&    \frac{ 2x \xi \Sigma}{a x}\{\Delta_\xi +\Delta_{\Sigma}\}  
                   \nonumber \\  
&+&  2Gm\bar{n}aQx\xi ^{2(1+\delta)} (1+\delta) M'_{\gamma(1+\delta),q}
 \Delta_\xi=0,    \nonumber \\  
&&~~~~~~~~~~~~~~~~~~~~~~~~~~({\rm 1st~moment})  
\label{24}  
\end{eqnarray}  
\begin{eqnarray}  
\lefteqn{  
         \frac{1}{a^2}\frac{\partial}{\partial t} [a^2 \xi \Pi]
          \{\Delta_\xi +\Delta_{\Pi}\}  
            }     \nonumber \\  
&&~~+  \frac{1}{a x^2}  
     \frac{\partial}{\partial x}  
        \left[  
        x^2 \xi \{ 
                  3\langle v \rangle\Pi
                     \{\Delta_\xi +\Delta_{\langle v \rangle}+\Delta_{\Pi}\}  
                  +s_\parallel
                     \{\Delta_\xi +\Delta_{s_\parallel}\}  
                 \}  
        \right]    
     \nonumber \\  
&&~~-
      \frac{4\xi}{a x} \{
                \langle v \rangle \Sigma
                   \{\Delta_\xi +\Delta_{\langle v \rangle}+\Delta_{\Sigma}\}  
                + s_\perp 
                   \{\Delta_\xi +\Delta_{s_\perp}\}  
                        \}  
                                  \nonumber \\  
&&~~+  4Gm\bar{n}aQ^* x  \langle v \rangle
           \xi ^{2(1+\delta)} 
      [(1+\delta) M'_{\gamma(1+\delta),q} \Delta_\xi + M_{\gamma(1+\delta)} 
\Delta_{\langle v \rangle}]=0,
                                   \nonumber \\  
&&~~~~~~~~~~~~~~~  
~~~~~~~~~~~~~({\rm 2nd~moment:contraction~ 1})  
\label{25}  
\end{eqnarray}  
\begin{eqnarray}  
\lefteqn{
      \frac{1}{a^2}\frac{\partial}{\partial t}  
      [a^2 \xi \Sigma \{\Delta_\xi +\Delta_{\Sigma}\}  ]
        } \nonumber \\  
&+&   \frac{1}{a x^4}  
      \frac{\partial}{\partial x}  
          \left[  
           x^4 \xi [
               \langle v \rangle \Sigma
                  \{\Delta_\xi +\Delta_{\langle v \rangle}+\Delta_{\Sigma}\}  
             + s_\perp 
                  \{\Delta_\xi +\Delta_{s_\perp}\}
                    ]
           \right] = 0.  
                                    \nonumber \\  
&&~~~~~~~~~~~~~~  
~~~~~~~~~~~({\rm 2nd~moment:contraction~2})  
\label{26}  
\end{eqnarray}  
Here it should be noted that in deriving the above equations
(\ref{23})-(\ref{26}) we neglect some terms in full equations because of 
the following two reasons; one of them is that some terms are high 
order ones in the strongly non-linear limit.
Another reason is that some terms are the higher order ones in the limit 
of small perturbation (larger than the first order perturbation). 
he ordering parameters in bothlimits are generally independent of each other.
However we consider the case that the higher order terms in the strongly 
non-linear limit are much smaller than the first order terms of the 
perturbation. Then we can derive the above equations by neglecting 
higher order terms in the nonlinear approximation.
This case means that we neglect the effects of the higher order terms in the 
strongly non-linear approximation in considering the linear perturbation.

RF expected the power law perturbation given by the following;
\begin{equation}
\Delta_\xi=\epsilon_\xi a^p x^q.  
\end{equation}

In this case, $M_{\gamma (1+\delta )}$ and $M'_{\gamma (1+\delta ),q}$
are given by 

\begin{eqnarray}  
M_{\gamma (1+\delta )}
 &=& \int \frac{\mu}{y^2}s^{-\gamma (1+\delta )}(1+y^{-\gamma (1+\delta )}) 
      d^3 y,\\
M'_{\gamma (1+\delta ),q} &=& \int \frac{\mu}{y^2}
          [s^{q-\gamma (1+\delta )}(1+y^{-\gamma (1+\delta ) })
         + s^{-\gamma (1+\delta )}(1+y^{q-\gamma (1+\delta )})]  d^3 y,\\
\mu &=& \frac{x^\alpha}{x}\frac{z^\alpha}{z},~~~
y = \frac{x_{31}}{x_{21}}=\frac{z}{x},~~~
 s=\frac{x_{23}}{x_{21}}=(1+y^2 -2y\mu)^{1/2}.
\end{eqnarray}  
  
The integrals should not diverge. Then,
$2-\gamma(1+\delta)>0$ must be satisfied for $y\rightarrow 0$ 
and $\gamma(1+\delta)>0$ for $y\rightarrow \infty$ in the $M$.
Furthermore,
 $2+q-\gamma(1+\delta)>0$ must be 
satisfied for $y\rightarrow 0$ and $q-\gamma(1+\delta)<0$ 
for $y\rightarrow \infty$ in the $M'$.
As a result, the following relations must be satisfied;
\begin{equation}
0< \gamma (1+\delta)< 2,~~~~~~\gamma (1+\delta )-2 < q < \gamma (1+\delta ).
\label{31}
\end{equation}
The four perturbation equations (\ref{23})-(\ref{26})
 are a little different from 
the equations that RF derived.
This is because 
in perturbating the three-body correlation term, 
RF devide the $\langle v_{21} \rangle$ into 
$\langle v_{23} \rangle+ \langle v_{31} \rangle$ artificially 
and perturbed $\langle v_{23} \rangle$ and $ \langle v_{31} \rangle$ 
independently, which is incorrect treatment.
Furthermore, although they obtained the $q$-dependent of $M'$,
they used the value of $M'$ only at $q=0$ in solving the 
perturbation equations.

Here we consider the following form of the perturbation.
We need to put the perturbations that do not diverge in the strongly 
non-linear limit $(x\rightarrow 0)$
and the linear limit $(x\rightarrow \infty)$.
Then we put the perturbation of the two-point spatial correlation function 
as follows;

\begin{equation}
\Delta_\xi=\epsilon_\xi a^p x^q,  \nonumber 
\label{32}
\end{equation}
\begin{equation}
q=|q|i,
\label{33}
\end{equation}
where
$|q|$ is the real number, that is, $q$ is the pure imaginary number 
while RF adopted the real number.
In this case, the perturbations do not diverge in the strongly 
non-linear limit and also in the linear limit.
If $q$ is real number as RF adopted, the perturbation diverges on some scales.
When $q$ is negative, the perturbation diverges in the non-linear limit
$(x\rightarrow 0)$. On the other hand, 
when $q$ is positive, the perturbation diverges in the linear limit
$(x\rightarrow \infty)$.
The perturbations of the other valuables also can be put in the same form. 
From the dimensional analysis, all perturbations must have the same 
value of the power ($q$ and $p$) as that of the two-point spatial correlation
function. When $q$ is pure imaginary number, $|s^q|=1$ and $|y^q|=1$.
And the following relations are satisfied;
\begin{equation}
|M'_{\gamma (1+\delta ),q}|\leq M'_{\gamma (1+\delta ),q=0}
=2M_{\gamma (1+\delta )}. 
\end{equation}
When $0<\gamma (1+\delta )<2$, both $M_{\gamma (1+\delta )}$ 
and $M'_{\gamma (1+\delta ),q}$ are finite for any value 
of pure imaginary number $q$. 

These perturbed BBGKY equations are not closed 
by themselves in general and higher moment equations are needed. 
But if the coefficient
of the perturbation of the skewness ($\Delta_{s_\parallel}$ and 
$\Delta_{s_\perp}$)
are equal to 0, 
the perturbation equations for the other valuables have 
no relation to the perturbation of the skewness.
In this case we can solve these perturbation equations independently 
from the higher moment equations.
As we can see from eqs.(\ref{25}) and (\ref{26}),
when the skewness of the background velocity field is equal to 0, 
the coefficient
of the perturbation of the skewness becomes 0.
Here we consider only this case that the skewness is equal to 0.
Then we can neglect the higher moment equations.
RF treated $A$ and $B$ as a constant value. So, the treatment of the 
perturbation of the skewness itself was mistaken.
But when the skewness is equal to 0, RF treatment happen to be 
correct fortunately.
In this case, the above equations are rewriten by putting the 
perturbations which have the power law given by eqs.(\ref{32})
 and (\ref{33}) as follows;

\begin{equation}
(p-hq)\Delta _\xi-h(3-\gamma +q)\Delta _{\langle v \rangle}=0,
\label{35}
\end{equation}

\begin{eqnarray}
\lefteqn{  
[2\{\gamma (1+\delta)-2+\sigma \}(1+2\delta)+q+2(1+\delta )Dkq]\Delta _\xi
         }\nonumber \\
&+&\{ 4-2\gamma (1+\delta )+q\}\Delta _{\Pi} -2\sigma \Delta _\Sigma =0,
\label{36}
\end{eqnarray}

\begin{eqnarray}
\lefteqn{  
[1-15h+(6+4\gamma )h(1+\delta )+4h\sigma +p-3hq 
         }\nonumber \\
&&-8h\frac{Q^*}{Q}\{\gamma (1+\delta )-2+\sigma \}(1+\delta)
-4h\frac{Q^*}{Q}(1+\delta )Dkq]\Delta _\xi  \nonumber \\
&&+[-3h\{ 5-2\gamma (1+\delta )+q\} 
+4h\sigma -4h\frac{Q^*}{Q}\{\gamma (1+\delta )-2+\sigma \}]
\Delta _{\langle v \rangle}  \nonumber \\
&&+[1-15h+(6+4\gamma )h(1+\delta )+p-3hq]
\Delta _{\Pi}+4h\sigma \Delta _{\Sigma}=0,
\label{37}
\end{eqnarray}

\begin{equation}
(1-h+6h\delta +p-hq)\Delta _\xi-h\{ 7-2\gamma (1+\delta )+q\}
\Delta _{\langle v \rangle}+(1-h+6h\delta +p-hq)\Delta _\Sigma =0,
\label{38}
\end{equation} 

where

\begin{equation}
D \equiv  \gamma (1+\delta )-2+\sigma ,~~~~~~
\sigma=\frac{\Sigma}{\Pi}.
\label{39}
\end{equation} 

It is difficult to treat exactly $M'$ as a function of $q$.
We approximate $M'$ by linear function of $q$. When $q=0$, $M'/M$
is equal to $2$.
Then we use the following approximation of $M'/M$
in the above equations (\ref{24}) and (\ref{25});

\begin{equation}
\frac{M'}{M} \equiv 2+kq,
\end{equation} 
where
\begin{equation}
k=\frac{ \int \frac{\mu}{y^2}
          [s^{-\gamma (1+\delta )}(1+y^{-\gamma (1+\delta ) })\log s
         + s^{-\gamma (1+\delta )}y^{-\gamma (1+\delta )} \log y]  d^3 y  }
{\int \frac{\mu}{y^2}s^{-\gamma (1+\delta )}(1+y^{-\gamma (1+\delta )}) d^3 y}.
\label{40}
\end{equation} 

 The integral 
$\int (\mu/y^2) s^{-\gamma (1+\delta )}(1+y^{-\gamma (1+\delta )}) d^3 y$
 is dominated arround the $s$ and $y$ $\sim 1$.
Arround the $s$ and $y$ $\sim 1$, $\log s$ and $\log y$ have the values of 
order 1. So, $k$ has the value of order 1.

Furthermore we use the following relation that can be derived from the 
first moment equation [eq.(\ref{13})];

\begin{equation}
4-2\gamma (1+\delta )-2\sigma 
+2Gm\bar{n}a^2 Qx^2 \frac{\xi ^{1+2\delta }}{\Pi}M_{\gamma (1+\delta )}=0.
\end{equation} 

These four equations [eqs.(\ref{35})-(\ref{38})]
 can be rewriten by using a matrix expression.

\begin{equation}
N_{ij}u_j=0,  \nonumber 
\end{equation} 
\begin{equation}
u_j=(\Delta_\xi,\Delta_{\langle v \rangle},\Delta_\Pi,\Delta_\Sigma ).
\end{equation} 

If there exist a non trivial solution, 
the determinant of the 
matrix $N_{ij}$ should be equal to 0.
Now we consider the zero skewness case, there is the relation between
the three-point spatial correlation function and the mean relative pecurier 
velocity by using eq.(\ref{20}). 
\begin{equation}
\delta =\frac{h-1}{6h}.
\label{44}
\end{equation} 
By using this relation, we can eliminate $\delta$ in 
eqs.(\ref{36})-(\ref{38}).
Furthermore, from the first moment equation(\ref{13}) and the second moment 
(contraction 1) equation (\ref{14}), we obtain the following relation;
\begin{equation}
(-h+1+6h\delta)\Pi +\{ 5-2\gamma (1+\delta )\}\frac{s_\parallel}{\dot{a}x}
-\frac{4s_\perp}{\dot{a}x}
-4Gm\bar{n}a^2 x^2 M_{\gamma (1+\delta )}h(Q^*-Q)=0.
\label{45}
\end{equation} 
As we can see, when the skewness is equal to 0, $Q^*-Q$ must be 0,
that is, $Q^*/Q$ is equal to 1.

Then the determinant of {\boldmath $N$}  is given by
\begin{equation}
\det \mbox{\boldmath $N$} =\frac{1}{3}(p-hq)^2 f(q,\gamma ',h,\sigma,kD), 
\end{equation} 
where

\begin{eqnarray}  
f(q,\gamma ',h,\sigma,kD) &=&\{9h + kD(7h-1)\}q^2   \nonumber \\   
&+&(4-2\gamma '+29h-10\gamma 'h-2\sigma +8h\sigma)q   \nonumber \\   
&+&kD(-3+21h-6\gamma 'h)q+12-6\gamma '-6\sigma,
\end{eqnarray} 
and
\begin{equation}
\gamma '\equiv \gamma (1+\delta). 
\end{equation}  
Here $f$ is the quadratic equation of $q$. Now we investigate whether 
the  equation $f=0$ has real solutions or not.
Since we do not know the value $\gamma '$, $h$, $\sigma$, and $kD$ in the
strongly non-linear regime, we treat these values as  parameters. 
Here we consider the allowed range of these parameter.
As we can see from  eq.(\ref{31}), the parameter $\gamma '$ must be satisfied 
$0<\gamma '<2$.
Since we have treated as $\zeta \gg \xi$ in the strongly non-linear regime, 
$\xi^{2(1+\delta )}$ should be higher order than $\xi$.
This results in that $2(1+\delta ) >1$ must be satisfied.
In this case, $h>\frac{1}{4}$ must be satisfied from eq.(\ref{44}).
YG showed the probable range of the mean relative peculier velocity 
and obtained that the mean relative physical pecurier velocity must 
have the value between 0 and the Hubble expansion velocity.
This means that the parameter $h$(relative velocity parameter) have the value 
between 0 and 1. So in this case, the parameter $h$ should be 
in the range $\frac{1}{4}<h<1$.
We do not know the range about the parameters $\sigma$ and $kD$.
But the parameter $\sigma$ sould have an order of 1.
Hence we investigate $\sigma$ in the range $\frac{1}{2}<\sigma<2$.
The parameter $D$ and $k$ also take an order of 1 as seen from the eqs.
(\ref{39}) and (\ref{40}), respectively.
So we investigate $kD$ in the range $-1<kD<1$.

In the above probable value of the parameters,
we can easily ascertain that 
$f=0$ has  real solutions. In other words,  $f=0$ does not have the
solutions of imaginaly number.
Since we consider the case that $q$ is imaginaly number,
$f=0$ can not be satisfied.
$p-hq$ should be 0 to satisfy that the determinant of 
the matrix {\boldmath $N$} is equal to 0. 
In this case, $p$ has the following value
\begin{equation}
p=hq=h|q|i.
\end{equation} 
This means that the perturbations do not grow.
And the solutions are stable.

Furthermore we consider the strict condition which determines the stability 
of the two-point spatial correlation function.
In the strongly non-linear regime, the mean relative peculier velocity 
has the value, $\langle v \rangle =-h\dot{a}x$ according to the process 
of clustering.
In this case, the scale of $a^h x$ for the two particles 
whose mean comoving distance is $x$
does not change as we can see from the following relation;

\begin{eqnarray}  
\frac{d}{dt}(a^h x)&=& a^{h-1} (a\dot{x}+h\dot{a}x)    \nonumber \\
 &=& a^{h-1} ( \langle v \rangle +h\dot{a}x )          \nonumber \\
 &=& 0 
\end{eqnarray} 
Then we should determine the stability of the two-point spatial correlation 
function at the fixed scale of $a^h x$.
The solutions of the perturbation are rewritten by
\begin{eqnarray} 
\Delta_\xi &=&\epsilon _\xi (a^h x)^q \nonumber \\
&=& \epsilon _\xi e^{i|q| log(a^h x)}
\label{49}
\end{eqnarray} 
This perturbation does not grow nor does it decay.
At the fixed scale of $a^hx$, the perturbation never even oscillate.  
This means that the perturbation is marginal stable. 
RF used the real number $q$ in investigating the behaviour of the perturbation.
As we can see from eq.(\ref{49}), the perturbation in the $p-hq=0$ mode 
 works well even when $q$ is real number. That is, the perturbation 
is marginal stable in this mode.
However there also exist a 'strange' growing mode 
for the real number of $q$ because $f=0$ is satisfied in this case.   
RF did not comment sufficiently about this 'strange' growing mode.

When $q$ is negative, the perturbation diverges in the non-linear limit
$(x\rightarrow 0)$. On the other hand, 
when $q$ is positive, the perturbation diverges in the linear limit
$(x\rightarrow \infty)$.
Then this perturbation is not adequate in investigating the local 
stability of the two-point spatial correlation function in the 
strongly nonlinear regime.

\section{RESULTS AND DISCUSSION}  
  
In this paper, we investigated the stability of the scale-invariant solutions 
of the cosmological BBGKY equations in the strongly nonlinear regime in the 
case that the skewness of the velocity field is equal to zero.
The reason why we consider the only case that the skewness vanishes 
is that perturbed BBGKY equations for $\xi$, $\langle v \rangle$, 
$\Pi$ and $\Sigma$ can be closed independently
of the higher moment perturbations.
When the power law perturbations are put in the solutions as RF put,
that is, when $q$ is real number,
the perturbation in the linear limit or the non-linear limit diverges.
When $q$ is negative, 
the perturbations diverge and do not work in the non-linear limit.
So $q$  should be positive.
On the other hand, when $q$ is positive, the perturbations work 
well in the non-linear regime. 
In the linear regime, however, those perturbations may 
diverge 
if the power law form of the pertubations are retained 
and so
the form of the perturbations should be changed in order not to 
diverge in the linear limit.
At this case, it is insufficient to solve the non-linear approximated 
equations because the informations about the evolutions on all 
scales are needed.
Then we investigated only the local stability of the non-linear regime.
In investigating the local stability of the two-point spatial 
correlation function in the strongly non-linear regime, 
we should put the perturbation whose value is zero or smaller value 
on the scales except for the strongly non-linear scales
than that on the strongly non-linear scalesthe investigating scale. 
That is, we should put the wave packet-like perturbation.
In order to put such a wave packet-like perturbation,
$q$ must be imaginaly number. In this case, we found that 
there is no unstable mode in the strongly non-linear regime.
It seems stable for any value of the power index of the two-point 
spatial correlation function in the strongly non-linear regime.
However we do not know whether a grobal unstability exist or not
because we consider only the local stability. 
It is certain that there is no local unstability in the 
strongly non-linear regime.
So, in the  strongly non-linear regime, the solutions is marginally stable,
and it does not seem that the power index of the two-point spatial
correlation function approaches to some stable point values.
The power index of the two-point spatial correlation function 
that was derived by DP, $\gamma=3(3+n)/(5+n)$, is not the special one
also in terms of the stability of the solution.
As a result, the argument of the stability does not determine the power 
index of the two-point spatial correlation function in the 
strongly non-linear regime.

The power index of the two-point spatial correlation function
is determined only by the clustering process, that is, the 
parameter $h$ if the self-similar solutions exist.
Hence it is very important to estimate the parameter $h$ and 
investigate whether the self-similar solutions exist or not in the 
general scale-invariant solutions which YG obtained.

\acknowledgements  
  
We are grateful to M.Nagashima for useful discussions. 
We would like to thank T.Tanaka for important comments.
And we would like to thank S.Ikeuchi and M.Sasaki for useful suggestions 
and continuous encouragement.  This work was supported in part by 
the Grant-in-Aid No.06640352 for the Scientific Research Fund from the 
Ministry of Education, Science and Culture of Japan.

\end{document}